\documentclass[12pt]{article}
\usepackage{amsmath}
\usepackage{amssymb}
\usepackage{amsfonts}
\usepackage{graphics}

\newcommand{\field}[1]{\mathbb{#1}}
\newcommand{\R}{\field{R}}

\title{An update on the classical and quantum harmonic oscillators on the sphere and the hyperbolic plane
in polar coordinates}

\author{C.\ Quesne\thanks{E-mail address: cquesne@ulb.ac.be} \\
{\small\sl Physique Nucl\'eaire Th\'eorique et Physique Math\'ematique, 
Universit\'e Libre de Bruxelles,} \\ 
{\small \sl Campus de la Plaine CP229, Boulevard~du Triomphe, B-1050
Brussels, Belgium}}
\date{ }
\begin{document}
\baselineskip=22pt plus 1pt minus 1pt
%%%%%%%%%%%%%%%%%%%%%%%%%%%%%%%%%%%%%%%%%%%%%%%%%%%%%%%%%%
\maketitle

\begin{abstract}
A simple derivation of the classical solutions of a nonlinear model describing a harmonic oscillator on the sphere and the hyperbolic plane is presented in polar coordinates. These solutions are then related to those in cartesian coordinates, whose form was previously guessed. In addition, the nature of the classical orthogonal polynomials entering the bound-state radial wavefunctions of the corresponding quantum model is identified.
\end{abstract}

\vspace{0.5cm}


 
\newpage
%
%========================================================================
%
\section{Introduction}

During many years, there has been a continuing interest for some generalizations \cite{carinena04a, carinena07a, ranada, cq} of a classical nonlinear oscillator \cite{mathews, lakshmanan}, which was introduced as a one-dimensional analogue of some quantum field theoretical models, and for the corresponding extensions \cite{carinena07a, ranada, carinena07b, carinena07c, midya, schulze12, schulze13} of its quantum version \cite{carinena04b, carinena07d}. Such a model is indeed an interesting example of a system with nonlinear oscillations with a frequency showing amplitude dependence. Furthermore, since it contains both a nonlinear potential and a position-dependent mass, it is amenable to applications in those areas wherein the harmonicity of vibrations breaks down, such as in high-energy molecular states, or wherein a position-dependent effective mass is a useful concept, such as in many condensed-matter systems or many-body problems.\par
%
%-----------------------------------------------------------------------------------------------------------
% 
In 2004, Cari\~nena, Ra\~nada, Santander, and Senthilvelan introduced a two-dimensional (and more generally $n$-dimensional) classical generalization \cite{carinena04a} of the one-dimensional model of \cite{mathews, lakshmanan}. They established that the nonlinearity parameter $\lambda$, entering the definition of the potential and the position-dependent mass, can be interpreted as $-\kappa$, where $\kappa$ is the curvature of the two-dimensional space, so that their model actually describes a harmonic oscillator on the sphere (for $\lambda = - \kappa < 0$) and on the hyperbolic plane (for $\lambda = - \kappa > 0$). They presented the solutions of the Euler-Lagrange equations in cartesian coordinates, showed that the system is superintegrable and that the Hamilton-Jacobi equations are separable in three different coordinate systems. Later on, the quantum version of this classical model was also exactly solved in the corresponding three coordinate systems, wherein the Schr\"odinger equation is separable \cite{carinena07b, carinena07c}.\par
%
%-----------------------------------------------------------------------------------------------------
%
In the present Letter, we deepen our understanding on these two-dimensional classical and quantum models by providing an update on their analysis in polar coordinates.\par
%
%------------------------------------------------------------------------------------------------------------------
%
In \cite{carinena04a}, the Euler-Lagrange equations in cartesian coordinates were so complicated that they could not be directly solved in a simple way. Some particular expressions were then assumed for the solutions and the undetermined parameters they contained were shown to satisfy some constraints by inserting such expressions in the equations. Here we plan to prove that, in contrast, the Euler-Lagrange equations in polar coordinates simplify considerably, so that their solutions can be systematically derived.\par
%
%-------------------------------------------------------------------------------------------------
%
{}Furthermore, in the solutions of the quantum model in polar coordinates \cite{carinena07c}, the precise nature of the classical orthogonal polynomials entering the bound-state radial wavefunctions was not determined. We will complete this analysis here, thereby extending to two dimensions a recent study \cite{schulze12}, wherein the quantum one-dimensional model of \cite{carinena04b} and \cite{carinena07d} was re-examined.\par
%
%===================================================================
%
\section{Solutions of the Euler-Lagrange equations in polar coordinates}

In cartesian coordinates $x$, $y$, the Lagrangian of \cite{carinena04a} can be written as
\begin{equation}
  L = \frac{1}{2} \frac{1}{1 + \lambda(x^2+y^2)} [\dot{x}^2 + \dot{y}^2 + \lambda (x\dot{y}-y\dot{x})^2]
  - \frac{1}{2} \frac{\alpha^2 (x^2+y^2)}{1 + \lambda(x^2+y^2)},
\end{equation}
where $\lambda$ may be positive or negative and $\alpha$ is some real constant that we may assume positive. In polar coordinates, it can be rewritten as
\begin{equation}
  L = \frac{1}{2} \left(\frac{\dot{r}^2}{1 + \lambda r^2} + \frac{J^2}{r^2}\right) - \frac{1}{2} \frac{\alpha^2 
  r^2}{1 + \lambda r^2}, 
\end{equation}
where the angular momentum $J = x\dot{y}-y\dot{x} = r^2 \dot{\varphi}$ is a constant of the motion.\par
%
%---------------------------------------------------------------------------------------------------------
%
Considering now the Euler-Lagrange equations, we get a single differential equation to solve, namely
\begin{equation}
  \ddot{r} - \frac{\lambda r}{1 + \lambda r^2} \dot{r}^2 + \frac{\alpha^2 r}{1 + \lambda r^2} - J^2
  \frac{1 + \lambda r^2}{r^3} = 0,  \label{eq:E-L}
\end{equation}
since the constancy of $J$ ensures that the other equation with respect to $\varphi$ is automatically satisfied. To solve (\ref{eq:E-L}), we proceed in two steps.\par
%
%-----------------------------------------------------------------------------------------------------------
% 
{}First, on setting $\dot{r} = p(r)$, we obtain a first-order equation for $p^2$,
\begin{equation}
  \frac{dp^2}{dr} - \frac{2\lambda r}{1 + \lambda r^2} p^2 + \frac{2\alpha^2 r}{1 + \lambda r^2}
  - 2J^2 \frac{1 + \lambda r^2}{r^3} = 0,
\end{equation}
whose general solution is given by
\begin{equation}
  p^2(r) = C (1 + \lambda r^2) - \frac{J^2}{r^2} + \frac{\alpha^2}{\lambda} - \lambda J^2,  \label{eq:p^2}
\end{equation}
in terms of some integration constant $C$. Second, from (\ref{eq:p^2}), we get the differential equation
\begin{equation}
  2dt = \frac{dr^2}{\sqrt{a + br^2 + cr^4}}, \qquad a = -J^2, \qquad b = C + \frac{\alpha^2}{\lambda} - 
  \lambda J^2, \qquad c = C\lambda,  \label{eq:diff-eqn}
\end{equation}
which can be easily integrated by taking into account the sign of the discriminant $\Delta = 4ac - b^2$ whenever $c \ne 0$ \cite{gradshteyn}. The solutions for $t = t(r^2)$ can then be inverted to yield $r^2 = r^2(t)$.\par
%
%------------------------------------------------------------------------------------------------------
%
{}Finally, the integration of the first-order differential equation $\dot{\varphi} = J/r^2(t)$ \cite{gradshteyn} provides the functions $\varphi = \varphi(t)$ for $J \ne 0$ in terms of some constant $K$ (since for $J=0$, $\varphi$ remains constant).\par
%
%----------------------------------------------------------------------------------------------------------
%
To write some physically-relevant results, it is worth observing that the value of the integration constant $C$ is directly related to the energy $E$ of the system. The latter is indeed given by
\begin{equation}
  E = \frac{1}{2} \frac{1}{1 + \lambda r^2} \left[\dot{r}^2 + \alpha^2 r^2 + \frac{J^2}{r^2}(1 + \lambda r^2)
  \right]  \label{eq:E}
\end{equation}
and insertion of (\ref{eq:p^2}) in (\ref{eq:E}) leads to
\begin{equation}
  E = \frac{1}{2} C + \frac{\alpha^2}{2\lambda} \qquad \text{or} \qquad C = 2E - \frac{\alpha^2}{\lambda}.
  \label{eq:E-C}
\end{equation}
On the other hand, Eq.~(\ref{eq:E}) can be rewritten as
\begin{equation}
  E = \frac{1}{2} \frac{\dot{r}^2}{1 + \lambda r^2} + V_{\rm eff}(r), \qquad V_{\rm eff}(r) = \frac{1}{2}
  \frac{\alpha^2 r^2}{1 + \lambda r^2} + \frac{J^2}{2r^2},
\end{equation}
where the constancy of $J$ allows us to group the term $J^2/(2r^2)$, coming from the kinetic energy, with the potential $V(r) = \alpha^2 r^2/[2(1 + \lambda r^2)]$ to define an effective potential $V_{\rm eff}(r)$.\par
%
%-----------------------------------------------------------------------------------------
%
The possible values of $E$, and consequently of $C$, are determined by the behaviour of $V_{\rm eff}(r)$, where for $\lambda > 0$, $r$ varies on the interval $(0, + \infty)$, while for $\lambda < 0$, it is restricted to $\left(0, 1/\sqrt{|\lambda|}\right)$. According to whether $J=0$ or $J\ne 0$, $V_{\rm eff}(r)$ goes to 0 or $+\infty$ for $r \to 0$. On the other hand, $V_{\rm eff}(r)$ goes to $\alpha^2/(2\lambda)$ for $r \to \infty$ if $\lambda > 0$ or to $+\infty$ for $r \to 1/\sqrt{|\lambda|}$ if $\lambda < 0$. Moreover, it can be easily shown that for $J\ne 0$, $V_{\rm eff}(r)$ has a minimum $V_{\rm eff, min} = \frac{1}{2} |J| (2\alpha - \lambda |J|)$ at $r_{\rm min} = \sqrt{|J|/(\alpha - \lambda |J|)} \in (0, +\infty)$ or $\left(0, 1/\sqrt{|\lambda|}\right)$ (according to which case applies). Note that in the $\lambda > 0$ case, such a minimum only exists for $J$ values such that $|J| < \alpha/\lambda$, thereby showing that bounded trajectories are restricted to low angular momentum values. It is worth pointing out that such a limitation on bounded motions for $\lambda > 0$ was not reported in \cite{carinena04a} and that for $J=0$, one may set $V_{\rm eff, min} = 0$. In Figs.~1 and 2, some examples are plotted for $\lambda > 0$ and $\lambda < 0$, respectively.\par
%
%--------------------------------------------------------------------------------------------------------------
%
\begin{figure}[h]
\begin{center}
\includegraphics{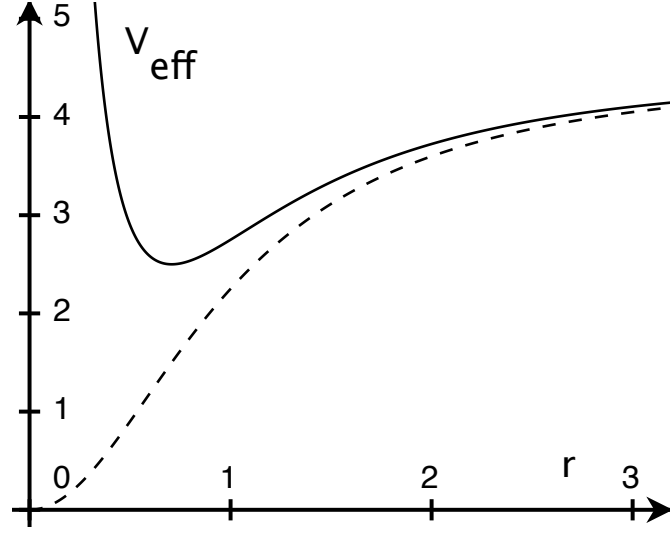}
\caption{Plot of $V_{\rm eff}(r)$, $\alpha=3$, $\lambda=1$, as a function of $r$ for $J=0$ (dashed line) and $J=1$ (solid line).}
\end{center}
\end{figure}
\par
%
%-------------------------------------------------------------------------------------------------------------------------------
%
\begin{figure}[h]
\begin{center}
\includegraphics{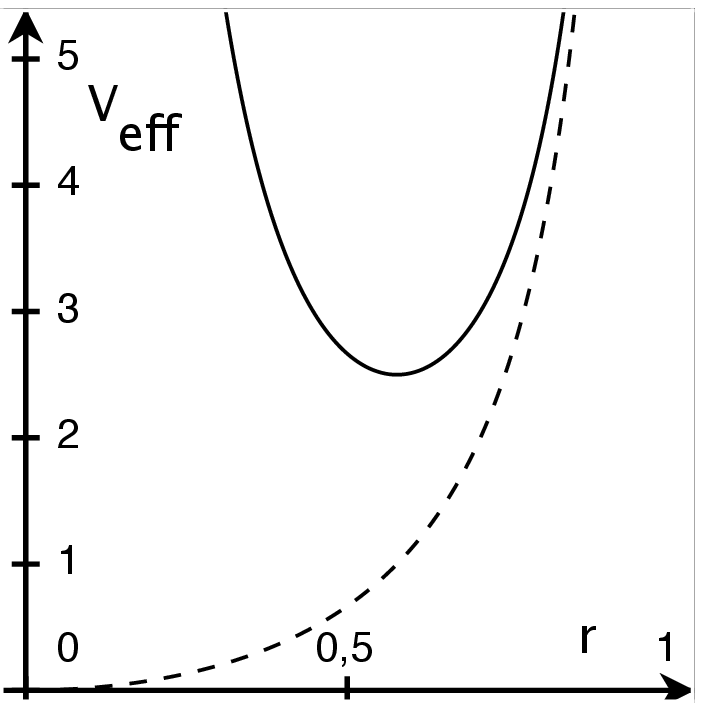}
\caption{Plot of $V_{\rm eff}(r)$, $\alpha=2$, $\lambda=-1$, as a function of $r$ for $J=0$ (dashed line) and $J=1$ (solid line).}
\end{center}
\end{figure}
\par
%
%-------------------------------------------------------------------------------------------------------------------------------
%
The restrictions on the constants $C$, $c$, and $\Delta = - (2E + \lambda J^2 - 2\alpha J)(2E + \lambda J^2 + 2\alpha J)$ of Eq.~(\ref{eq:diff-eqn}) for each energy domain are given by
\begin{equation}
\begin{split}
  \text{(i)} & \quad \text{If $\lambda>0$ and $V_{\rm eff,min} < E < \alpha^2/(2\lambda)$ or if $\lambda<0$
     and $V_{\rm eff,min} < E < +\infty$,} \\
  & \quad \text{then $- (\alpha - \lambda |J|)^2/\lambda < C < 0$, $c<0$, and $\Delta<0$;} \\
  \text{(ii)} & \quad \text{If $\lambda>0$ and $E = \alpha^2/(2\lambda)$, then $C=0$ and $c=0$;} \\
  \text{(iii)} & \quad \text{If $\lambda>0$ and $\alpha^2/(2\lambda) < E < +\infty$, then $0<C<+\infty$, 
     $c>0$, and $\Delta<0$.}
\end{split}
\end{equation}
\par
%
%------------------------------------------------------------------------------------------------------------
%
{}For $\lambda>0$ and $V_{\rm eff,min} < E < \alpha^2/(2\lambda)$ or $\lambda<0$, the complete solution is given by
\begin{equation}
\begin{split}
  & r^2 = A \sin(2\omega t + \phi) + B, \quad B-A \le r^2 \le B+A, \\
  & A = \frac{1}{2|\lambda|\omega^2} \sqrt{[(\alpha - \lambda J)^2 - \omega^2][(\alpha + \lambda J)^2 
      - \omega^2]}, \\
  & B = \frac{\alpha^2 - \lambda^2 J^2 - \omega^2}{2\lambda\omega^2}, \quad \phi \in [0, 2\pi), \\
  & \tan(\varphi - K) = \frac{\omega}{J} \left[B \tan\left(\omega t + \frac{\phi}{2}\right) + A\right] \quad 
      \text{if $J\ne 0$}, \\
  & \varphi = K \quad \text{if $J=0$},
\end{split}  \label{eq:sol-1}
\end{equation}
and describes bounded trajectories. As previously observed, such a solution only exists for $|J| < \alpha/\lambda$ whenever $\lambda>0$.\par
%
%----------------------------------------------------------------------------------------------
%
{}For $\lambda>0$ and $\alpha^2/(2\lambda) < E < +\infty$, the trajectories are unbounded and characterized by
\begin{equation}
\begin{split}
  & r^2 = A \cosh(2\omega t + \phi) + B, \quad A+B \le r^2 < +\infty, \\
  & A = \frac{1}{2\lambda\omega^2} \sqrt{[(\alpha - \lambda J)^2 + \omega^2][(\alpha + \lambda J)^2 
      + \omega^2]}, \quad B = - \frac{\alpha^2 - \lambda^2 J^2 + \omega^2}{2\lambda\omega^2}, \quad
      \phi \in \R, \\
  & \tan(\varphi - K) = \frac{\omega}{J} (A-B) \tanh\left(\omega t + \frac{\phi}{2}\right) \quad 
      \text{if $J\ne 0$}, \\
  & \varphi = K \quad \text{if $J=0$}.
\end{split}  \label{eq:sol-2}
\end{equation}
\par
%
%----------------------------------------------------------------------------------------------
%
{}Finally, for $\lambda>0$ and $E = \alpha^2/(2\lambda)$, we get a limiting unbounded trajectory, specified by
\begin{equation}
\begin{split}
  & r^2 = (At + \phi)^2 + B, \quad B \le r^2 < +\infty, \\
  & A = \sqrt{\frac{1}{\lambda}(\alpha^2 - \lambda^2 J^2)}, \quad B = \frac{\lambda J^2}{\alpha^2 -
      \lambda^2 J^2}, \quad \phi \in \R, \\
  & \tan(\varphi - K) = \frac{A}{J} (At + \phi) \quad 
      \text{if $J\ne 0$}, \\
  & \varphi = K \quad \text{if $J=0$}.
\end{split}  \label{eq:sol-3}
\end{equation}
\par
%
%----------------------------------------------------------------------------------------------
%
In Eqs.~(\ref{eq:sol-1}) and (\ref{eq:sol-2}), the parameter $\omega$ is defined by $\omega = \sqrt{|c|}$, so that on using Eqs.~(\ref{eq:diff-eqn}) and (\ref{eq:E-C}), it can be related to $C$ and finally to $E$. As a result, the energy can be written as
\begin{equation}
  E = \frac{\alpha^2 - \omega^2}{2\lambda} \qquad \text{or} \qquad  E = \frac{\alpha^2 + \omega^2}
  {2\lambda}  \label{eq:E-omega}
\end{equation}
for bounded or unbounded motions, respectively. It is also worth observing that in the $J=0$ special case, Eqs.~(\ref{eq:sol-1}), (\ref{eq:sol-2}), and (\ref{eq:sol-3}) lead to the relations $A=B$, $A=-B$, and $B=0$, respectively.\par
%
%------------------------------------------------------------------------------------------------------------
%
Having completed the solution of the Euler-Lagrange equations in polar coordinates, we may connect it with that in cartesian coordinates presented in \cite{carinena04a}. This is the purpose of the next section.\par
%
%============================================================
%
\section{Connection with the solutions in cartesian coodinates}

In the cases of bounded, unbounded, and limiting unbounded motions, Cari\~nena et al.\ proposed the following solutions \cite{carinena04a}
\begin{equation}
  x = A_1 \sin(\bar{\omega}t + \phi_1), \qquad y = A_2 \sin(\bar{\omega} t + \phi_2),  \label{eq:C-sol-1}
\end{equation}
\begin{equation}
  x = A_1 \sinh(\bar{\omega}t + \phi_1), \qquad y = A_2 \sinh(\bar{\omega} t + \phi_2),  \label{eq:C-sol-2}
\end{equation}
\begin{equation}
  x = A_1 t + B_1, \qquad y = A_2 t + B_2,  \label{eq:C-sol-3}
\end{equation}
respectively. The parameters $A_1$, $A_2$, $\phi_1$, $\phi_2$ of (\ref{eq:C-sol-1}) and (\ref{eq:C-sol-2}) and $A_1$, $A_2$, $B_1$, $B_2$ of (\ref{eq:C-sol-3}) are constrained by the following relations:
\begin{equation}
\begin{split}
  & \alpha^2 = M \bar{\omega}^2, \quad M = 1 + \lambda P_{\rm e}, \quad P_{\rm e} = A_1^2 + A_2^2
      + \lambda A_1^2 A_2^2 \sin^2 \phi_{12}, \quad \phi_{12} = \phi_1 - \phi_2, \\
  & J = \bar{\omega} A_1 A_2 \sin \phi_{12}, \quad E = \frac{1}{2} \bar{\omega}^2 (A_1^2 + A_2^2
      + \lambda A_1^2 A_2^2 \sin^2 \phi_{12}) = \frac{1}{2} \alpha^2 \frac{P_{\rm e}}{1+\lambda P_{\rm e}},
\end{split} \label{eq:C-rel-1}
\end{equation}
\begin{equation}
\begin{split}
  & \alpha^2 = M \bar{\omega}^2, \quad M = - 1 + \lambda P_{\rm h}, \quad P_{\rm h} = A_1^2 + A_2^2
      + \lambda A_1^2 A_2^2 \sinh^2 \phi_{12}, \quad \phi_{12} = \phi_1 - \phi_2, \\
  & J = \bar{\omega} A_1 A_2 \sinh \phi_{12}, \quad E = \frac{1}{2} \bar{\omega}^2 (A_1^2 + A_2^2
      + \lambda A_1^2 A_2^2 \sinh^2 \phi_{12}) = \frac{1}{2} \alpha^2 \frac{P_{\rm h}}
      {\lambda P_{\rm h} - 1},
\end{split}
\end{equation}
and
\begin{equation}
\begin{split}
  & \alpha^2 = \lambda P_{\rm L}, \quad P_{\rm L} = A_1^2 + A_2^2 + \lambda (A_2 B_1 - A_1 B_2)^2, \\
  & J = A_2 B_1 - A_1 B_2, \quad E = \frac{1}{2} [A_1^2 + A_2^2 + \lambda (A_2 B_1 - A_1 B_2)^2]
      = \frac{\alpha^2}{2\lambda},
\end{split}
\end{equation}
respectively.\par
%
%---------------------------------------------------------------------------------------------------------
%
Let us first compare the solution in polar coordinates for bounded trajectories with $J\ne0$, given in (\ref{eq:sol-1}) and (\ref{eq:E-omega}), with the corresponding one in cartesian coordinates, expressed in (\ref{eq:C-sol-1}) and (\ref{eq:C-rel-1}). To start with, direct comparison of the results obtained for the energy in (\ref{eq:E-omega}) and (\ref{eq:C-rel-1}) yields $\omega = \bar{\omega}$ for the angular frequency of the motion, as it should be. There remains in each case four constant parameters $A$, $B$, $\phi$, $K$ or $A_1$, $A_2$, $\phi_1$, $\phi_2$, related to the initial position and velocity. The connection between the two approaches will be established by expressing the former in terms of the latter.\par
%
%------------------------------------------------------------------------------------------------------------------
%
{}From the equation derived from $r^2 = x^2 + y^2$, we get
\begin{equation}
\begin{split}
  & B = \tfrac{1}{2} (A_1^2 + A_2^2), \\
  & A \cos\phi = \tfrac{1}{2} (A_1^2 \sin 2\phi_1 + A_2^2 \sin 2\phi_2), \quad
       A \sin\phi = - \tfrac{1}{2} (A_1^2 \cos 2\phi_1 + A_2^2 \cos 2\phi_2),
\end{split}  \label{eq:rel-1}
\end{equation}
which also yields
\begin{equation}
\begin{split}
  & A = \tfrac{1}{2} (A_1^4 + A_2^4 + 2 A_1^2 A_2^2 \cos 2\phi_{12})^{1/2}, \\
  & \tan\phi = - \frac{A_1^2 \cos 2\phi_1 + A_2^2 \cos 2\phi_2}{A_1^2 \sin 2\phi_1 + A_2^2 \sin 2\phi_2}.
\end{split}  \label{eq:rel-2}
\end{equation}
On the other hand, from $\tan(\varphi - K) = (\tan\varphi - \tan K)/(1 + \tan\varphi \tan K)$ and $\tan\varphi = y/x$, we arrive at the following constraint
\begin{equation}
\begin{split}
  & [A_2 \sin\phi_2 - A_1 \tan K \sin\phi_1 + \tan \omega t (A_2 \cos\phi_2 - A_1 \tan K \cos\phi_1)] \\
  & \quad \times\left(1 - \tan \omega t \tan \frac{\phi}{2}\right) \\
  & = \frac{\omega}{J} [A_1 \sin\phi_1 + A_2 \tan K \sin\phi_2 + \tan \omega t (A_1 \cos\phi_1 + A_2 
       \tan K \cos\phi_2)] \\
  & \quad {}\times \left[B \tan \frac{\phi}{2} + A + \tan \omega t \left(B - A \tan \frac{\phi}{2}\right)\right],
\end{split}
\end{equation}
leading to three conditions on the coefficients of 1, $\tan \omega t$, and $\tan^2 \omega t$. Such restrictions have been shown to be equivalent to a single one,
\begin{equation}
  \tan K = \frac{A_2 \sin\phi_2 - \frac{\omega}{J} A_1 \sin\phi_1 \left(B \tan \frac{\phi}{2} + A\right)}
  {A_1 \sin\phi_1 + \frac{\omega}{J} A_2 \sin\phi_2 \left(B \tan \frac{\phi}{2} + A\right)},  \label{eq:rel-3}
\end{equation}
expressing $\tan K$ in terms of $A_1$, $A_2$, $\phi_1$, $\phi_2$ after taking Eqs.~(\ref{eq:rel-1}) and (\ref{eq:rel-2}) into account.\par
%
%--------------------------------------------------------------------------------------------
%
By proceeding in a similar way in the cases of unbounded and limiting unbounded trajectories with $J\ne0$, we arrive at the relations
\begin{equation}
\begin{split}
  & \omega = \bar{\omega}, \quad B = - \tfrac{1}{2} (A_1^2 + A_2^2), \\
  & A \cosh\phi = \tfrac{1}{2} (A_1^2 \cosh 2\phi_1 + A_2^2 \cosh 2\phi_2), \quad
       A \sinh\phi = \tfrac{1}{2} (A_1^2 \sinh 2\phi_1 + A_2^2 \sinh 2\phi_2), \\
  & A = \tfrac{1}{2} (A_1^4 + A_2^4 + 2 A_1^2 A_2^2 \cosh 2\phi_{12})^{1/2}, \\
  & \tanh\phi = \frac{A_1^2 \sinh 2\phi_1 + A_2^2 \sinh 2\phi_2}{A_1^2 \cosh 2\phi_1 
       + A_2^2 \cosh 2\phi_2}, \\
  & \tan K = \frac{A_2 \sinh\phi_2 - \frac{\omega}{J} (A-B) A_1 \sinh\phi_1 \tanh \frac{\phi}{2}}
        {A_1 \sinh\phi_1 + \frac{\omega}{J} (A-B) A_2 \sinh\phi_2 \tanh \frac{\phi}{2}},
\end{split}  
\end{equation}
and
\begin{equation}
  A = (A_1^2 + A_2^2)^{1/2}, \quad B = \frac{(A_1B_2 - A_2B_1)^2}{A_1^2 + A_2^2}, \quad
  \phi = \frac{A_1B_1 + A_2B_2}{\sqrt{A_1^2 + A_2^2}}, \quad \tan K = - \frac{A_1}{A_2},
\end{equation}
respectively.\par
%
%--------------------------------------------------------------------------------------------------------
%
The $J=0$ case has not been specifically discussed in \cite{carinena04a}. From (\ref{eq:C-sol-1}) and (\ref{eq:C-rel-1}), however, one can see that it corresponds to $\sin \phi_{12} = 0$ or $\phi_{12} = k \pi$, which implies that $y = \pm A_2 x/A_1$. Equations (\ref{eq:rel-1}) and (\ref{eq:rel-2}) then lead to $A = B = \frac{1}{2}(A_1^2 + A_2^2)$ and $\tan \phi = - \cot 2\phi_1$, while Eq.~(\ref{eq:rel-3}) is replaced by $\tan \varphi = K = \pm A_2/A_1$. The correspondence for the remaining two types of trajectories can be easily established in a similar way.\par
%
%=========================================================
%
\section{Bound-state solutions of the Schr\"odinger equation in polar coordinates}

As shown in one dimension \cite{carinena04b, carinena07d}, the quantum version of the nonlinear oscillator of \cite{mathews, lakshmanan} is exactly solvable for a $\lambda$-dependent potential parameter $\alpha^2 = \beta (\beta+\lambda)$. The same is true for its two-dimensional generalization \cite{carinena07c}, whose Schr\"odinger equation reads
\begin{equation}
  \left((1+\lambda r^2) \frac{\partial}{\partial r^2} + (1+2\lambda r^2) \frac{1}{r} \frac{\partial}{\partial r}
  + \frac{1}{r^2} \frac{\partial^2}{\partial\varphi^2} - \frac{\beta(\beta+\lambda) r^2}{1+\lambda r^2}
  + 2E\right) \Psi(r,\varphi) = 0  \label{eq:S-E}
\end{equation}
in units wherein $\hbar=1$.\par
%
%----------------------------------------------------------------------------------------------
%
After separating the variables $r$ and $\varphi$ by setting
\begin{equation}
  \Psi(r,\varphi) = R(r) \frac{e^{{\rm i}m\varphi}}{\sqrt{2\pi}},
\end{equation}
where the angular momentum quantum number $m$ may be any positive or negative integer or zero, we arrive at the differential equation
\begin{equation}
  r^2 (1+\lambda r^2) R'' + r (1+2\lambda r^2) R' + \left(- \frac{\beta(\beta+\lambda) r^4}{1+\lambda r^2}
  + 2Er^2 - m^2\right) R = 0
\end{equation}
for the radial wavefunction $R(r)$ (with a prime denoting derivation with respect to $r$).\par
%
%----------------------------------------------------------------------------------------------------------------------
%
Let us now assume for $R(r)$ the following form
\begin{equation}
  R(r) = (1+\lambda r^2)^{-\beta/(2\lambda)} r^{|m|} f(r),  \label{eq:R}
\end{equation}
where $f(r)$ is some polynomial in $r$, satisfying the differential equation
\begin{equation}
  r (1+\lambda r^2) f'' + \{2|m| + 1 + 2[(|m|+1)\lambda - \beta] r^2\} f' + r [|m|(|m|+1)\lambda - 2 (|m|+1)
  \beta + 2E] f = 0.
\end{equation}
The expression (\ref{eq:R}) for $R(r)$ both eliminates the asymptotic behaviour, given by $(1+\lambda r^2)^{-\beta/(2\lambda)}$ for $r \to \infty$ if $\lambda > 0$ and for $r \to 1/\sqrt{|\lambda|}$ for $\lambda < 0$, and ensures that the radial wavefunction is finite for $r \to 0$.\par
%
%--------------------------------------------------------------------------------------------------------------
%
It only remains to change the variable $r$ into $t=1+2\lambda r^2$ to convert $f(r)$ into a function $g(t)$ fulfilling a differential equation (with a dot denoting derivation with respect to $t$)
\begin{equation}
\begin{split}
  & (1-t^2) \ddot{g} - \frac{1}{2\lambda} \{(2|m|+1)\lambda + 2\beta + [(2|m|+3)\lambda - 2\beta] t\} 
       \dot{g} \\
  & \quad {}+ \left[- \frac{1}{4} |m|(|m|+1) + \frac{(|m|+1)\beta - E}{2\lambda}\right] g = 0
\end{split}
\end{equation}
that we recognize as that satisfied by Jacobi polynomials $P^{(a,b)}_{n_r}(t)$, $n_r = 0$, 1, 2,~\ldots, namely \cite{gradshteyn}
\begin{equation}
  \left\{(1-t^2) \frac{d^2}{dt^2} + [b-a - (a+b+2)t] \frac{d}{dt} + n_r (n_r + a + b + 1)\right\}
  P^{(a,b)}_{n_r}(t) = 0.
\end{equation}
Here $a=|m|$, $b= -\frac{\beta}{\lambda} - \frac{1}{2}$, and the energy eigenvalues are found to be $E = - n_r [2\lambda(n_r+1) + (2|m|-1)\lambda - 2\beta] + (|m|+1) \left(\beta - \frac{\lambda}{2} |m|\right)$.\par
%
%---------------------------------------------------------------------------------------------------
%
The final result can be written as
\begin{equation}
\begin{split}
  & R_{n_r,|m|}(r) \propto (1+\lambda r^2)^{-\beta/(2\lambda)} r^{|m|} P^{\left(|m|, -\frac{\beta}{\lambda}
      - \frac{1}{2}\right)}_{n_r}(1+2\lambda r^2), \\
  & E_n = (n+1) \left(- \frac{\lambda}{2} n + \beta\right), \quad n = 2n_r + |m|.
\end{split}
\end{equation}
Here the range of $n$ values is determined from the normalizability of the radial wavefunction with respect to the measure $(1+\lambda r^2)^{-1/2} r dr$ and is given by
\begin{equation}
  n = \begin{cases}
     0, 1, 2, \ldots & \text{if $\lambda<0$}, \\
     0, 1, 2, \ldots, n_{\rm max}, \quad \frac{\beta}{\lambda} - \frac{3}{2} \le n_{\rm max} < \frac{\beta}
         {\lambda} - \frac{1}{2} & \text{if $\lambda>0$}.
  \end{cases}
\end{equation}
\par
%
%================================================================
%
\section{Conclusion}

In this Letter, we have proved that the Euler-Lagrange equations for the harmonic oscillator on the sphere and the hyperbolic plane, introduced by Cari\~nena et al., can be easily solved in polar coordinates and we have established the relation between our solution and that in cartesian coordinates presented by these authors. Furthermore, we have pointed out the existence of a restriction on the angular momentum values compatible with bounded motions on the hyperbolic plane, which was not reported before.\par
%
%---------------------------------------------------------------------------------------------------------------------
%
We have also demonstrated that the bound-state radial wavefunctions of the corresponding quantum problem can be written in terms of Jacobi polynomials. The result may be compared with the Laguerre polynomials entering the wavefunctions of the harmonic oscillator on the plane, as well as with the Gegenbauer polynomials making their appearance \cite{schulze12} in the one-dimensional quantum problem of \cite{carinena04b, carinena07d}.\par
%
%-------------------------------------------------------------------------------------------------------------------
%
As a final point, it is worth observing that the $\lambda$-deformed Hermite polynomials, shown to occur in the remaining two coordinate systems wherein the Schr\"odinger equation (\ref{eq:S-E}) is separable \cite{carinena07b}, can be identified as classical Gegenbauer polynomials too.\par
%
%==============================================================
%

\end{document}